\documentclass[aps,prl,twocolumn,showpacs,amsmath,amsmath,amssymb]{revtex4-1}

\usepackage{amsmath,amssymb,graphicx,bm,psfrag,bbold,float,color}

\usepackage{graphicx,physics,hyperref,appendix}

\usepackage[mathscr]{eucal}

\DeclareMathAlphabet{\mathpzc}{OT1}{pzc}{m}{it}

\begin{document}

\title{Charge-Qubit-Resonator-Interface-Based Nonlinear Circuit QED}

\author{Deshui Yu$^{1}$, Leong Chuan Kwek$^{1,2,3,4}$, Luigi Amico$^{1,5,6}$, \& Rainer Dumke$^{1,7}$}

\address{\mbox{$^{1}$Centre for Quantum Technologies, National University of Singapore, 3 Science Drive 2, Singapore 117543, Singapore}}

\address{$^{2}$Institute of Advanced Studies, Nanyang Technological University, 60 Nanyang View, Singapore 639673, Singapore}

\address{$^{3}$National Institute of Education, Nanyang Technological University, 1 Nanyang Walk, Singapore 637616, Singapore}

\address{$^{4}$MajuLab, CNRS-UNS-NUS-NTU International Joint Research Unit, UMI 3654, Singapore}

\address{\mbox{$^{5}$CNR-MATIS-IMM \& Dipartimento di Fisica e Astronomia, Universit\'a Catania, Via S. Soa 64, 95127 Catania, Italy}}

\address{$^{6}$INFN Laboratori Nazionali del Sud, Via Santa Sofia 62, I-95123 Catania, Italy}

\address{$^{7}$Division of Physics and Applied Physics, Nanyang Technological University, 21 Nanyang Link, Singapore 637371, Singapore}

\begin{abstract}
We explore applications of nonlinear circuit QED with a charge qubit inductively coupled to a microwave LC resonator in the photonic engineering and ultrastrong-coupling multiphoton quantum optics. Simply sweeping the gate-voltage bias achieves arbitrary Fock-state pulsed maser, where the single qubit plays the role of artificial gain medium. Resonantly pumping the parametric qubit-resonator interface leads to the squeezing of resonator field, which is utilizable to the quantum-limited microwave amplification. Moreover, upwards and downwards multiphoton quantum jumps may be observed in the steady state of the driving-free system.
\end{abstract}


\maketitle

\textit{Introduction.} Owing to properties of rapid operation, flexibility, and scalability, superconducting quantum circuits have become the most promising candidate for realizing quantum computation~\cite{RMP:Xiang2013}. However, the current practical execution is significantly limited by their short energy-relaxation and dephasing times (tens of $\mu$s~\cite{Science:Wang2016}). Feedback control may persist the Rabi oscillation of superconducting qubit infinitely~\cite{Nature:Vijay2012} and efficiently suppresses the low-frequency $1/f$ fluctuation in the circuit~\cite{NJP:Yu2018,PRA:Yu2018}. In addition, hybrid schemes composed of superconducting circuits and neutral atoms potentially implements the quantum-state transfer between rapid quantum processor and long-term memory~\cite{PRA:Yu2016-1,PRA:Yu2016-2,SciRep:Yu2016,QST:Yu2017,ProcSPIE:Hufnagel2017}.

Superconducting circuits also provide a vivid platform for exploring fundamental principles of the matter-light interaction, especially, in the ultrastrong-coupling regime where the conventional cavity QED system hardly accesses~\cite{NatPhys:Niemczyk2010,NatPhys:Yoshihara2017,PRA:Yu2017}. Multiple unique features have been recognized for circuit QED. Superconducting microwave resonators possess a quality factor much larger than high-finesse optical cavities~\cite{PRA:Blais2004}, resulting in a longer lifetime of intraresonator photons. The capacitive/inductive coupling between artificial atoms and resonator or between arbitrary two superconducting qubits in a many-body system may be designed deliberately and adjusted simply via tuning external voltage, current, or flux biases~\cite{PRL:Xu2018}, leading to flexible and engineerable networks~\cite{NatPhys:Houck2012}. Moreover, the nonlinear dispersive artificial-atom-resonator interaction overcomes the weak-response obstacle in the quantum measurement and efficiently enhances the readout fidelity~\cite{APL:Yamamoto2008,APL:Mutus2013}.

In this work, we theoretically investigate potential applications of a circuit-QED structure, where a charge qubit nonlinearly interacts with a resonator, in aspects of preparing the photon-number-state microwave light, squeezing radiation, and quantum jump process. We find that arbitrary Fock-state resonator field may be produced with a high fidelity by simply sweeping gate-charge and external flux biases. The resonant two-photon qubit-resonator coupling leads to the generation of squeezed radiation, suppressing the quantum fluctuation in measurement. In addition, multiphoton quantum jumps are identified in the steady state of the system. This nonlinear circuit QED paves the way for studying multiphoton quantum optics and quantum state engineering~\cite{PhysRep:DellAnno2006}.



\begin{figure}[b]
\includegraphics[width=8.5cm]{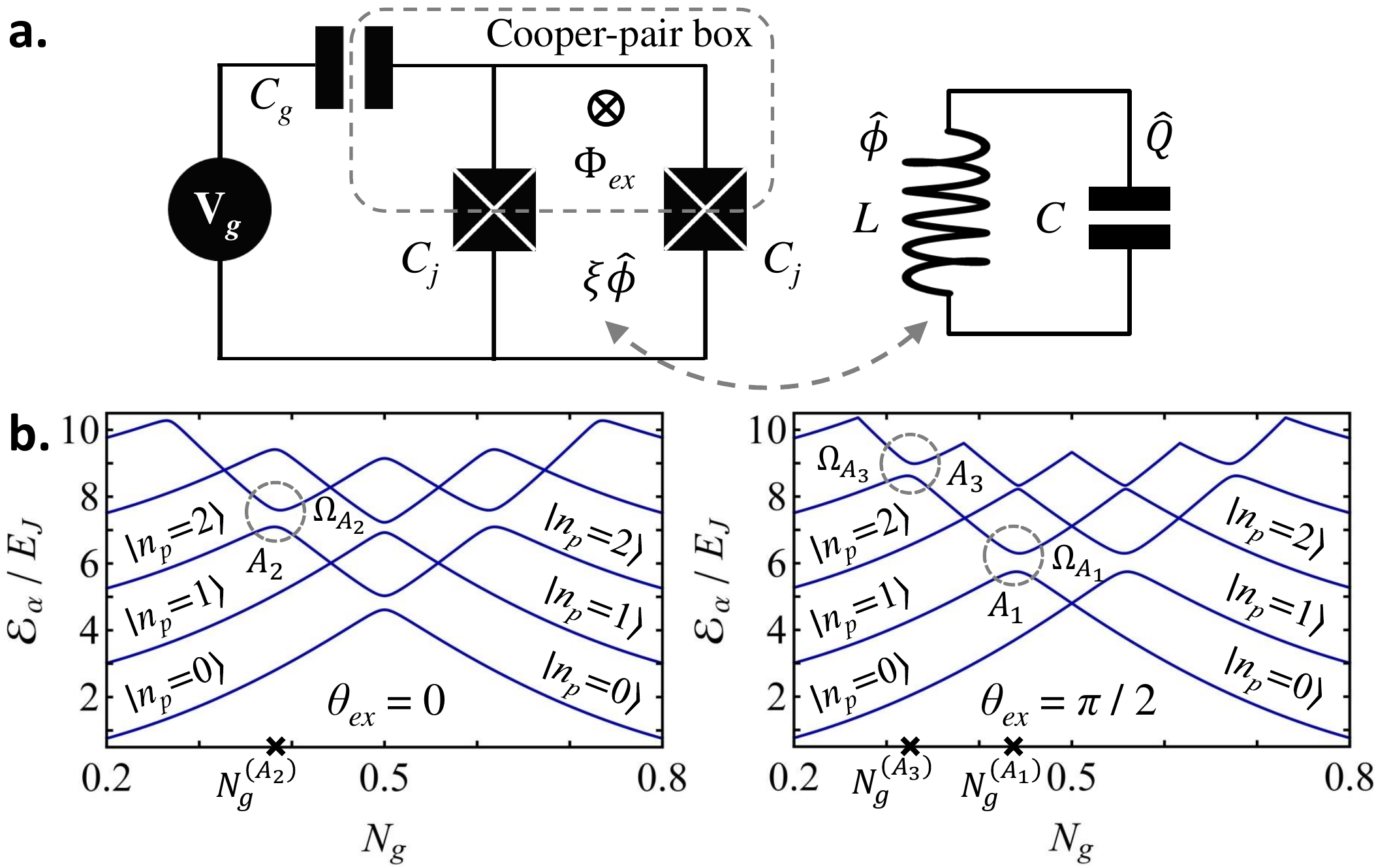}\\
\caption{(Color online) (a) Schematics of nonlinear charge-qubit-resonator interface. The charge qubit is composed of a pair of Josephson junctions with the self-capacitance $C_{j}$. A voltage source $V_{g}$ biases the Cooper-pair box via a gate capacitor $C_{g}$. An external flux $\Phi_{ex}$ is applied to tune the Josephson energy $E_{J}$ of Cooper pairs. The LC resonator consists of an inductor $L$ and a capacitor $C$. A portion of the flux from $L$, i.e., $\xi\hat{\phi}$, threads the charge-qubit loop, leading to the nonlinear inductive coupling. (b) Dependence of energy spectrum $\mathcal{E}_{\alpha}$ on $N_{g}$ with $\xi=1$. Three anticrossings, which are shown in circles and labeled by $A_{i=1,2,3}$, occur at $N^{(A_{i=1,2,3})}_{g}$. The corresponding energy gaps are $\hbar\Omega_{A_{i=1,2,3}}$.}\label{Fig1}
\end{figure}

\textit{Physical model.} We consider the nonlinear interface between a Cooper-pair box and an LC resonator as shown in Fig.~\ref{Fig1}(a). The box, which is biased by a voltage source $V_{g}$ via a gate capacitor $C_{g}=300$ aF, is connected to the Cooper-pair reservoir through two quasi-identical Josephson junctions with self-capacitance $C_{j}=50$ aF and Josephson energy $\frac{E_{J}}{2\hbar}=2\pi\times10$ GHz. The tunneling rate of Cooper pairs between the box and the reservoir is tuned by an external flux $\Phi_{ex}$~\cite{Nature:Makhlin1999}. The Cooper-pair box is characterized by the charge-number operator $\hat{N}$ and the phase difference $\hat{\delta}$ across
the Josephson junction with the commutator of $[\hat{\delta},\hat{N}]=i$. The microwave LC resonator consists of an inductor $L=100$ nH and a capacitor $C=500$ aF, resulting in the resonator frequency $\omega_{0}=\frac{1}{\sqrt{LC}}=2\pi\times22.5$ GHz. We use the operators $\hat{\phi}$ and $\hat{Q}$ to denote respectively the magnetic flux of $L$ and the charge on $C$ and have the commutator $[\hat{\phi},\hat{Q}]=i\hbar$. A portion of $\hat{\phi}$, measured by $\xi<1$, threads through the Cooper-pair-box loop, leading to the nonlinear inductive coupling. The whole system operates at the temperature $T=20$ mK with the thermal fluctuation $2\pi\times0.4$ GHz.

The coherent Cooper-pair-box-resonator interplay is governed by the Hamiltonian
\begin{eqnarray}\label{H}
\nonumber\hat{H}&=&E_{C}(\hat{N}-N_{g})^{2}+\hbar\omega_{0}\hat{a}^{\dag}\hat{a}\\
&&-E_{J}\cos[\theta_{ex}+\theta_{L}(\hat{a}^{\dag}+\hat{a})]\cos\hat{\delta},
\end{eqnarray}
where we have defined the charging energy $E_{C}=\frac{(2e)^{2}}{2(C_{g}+2C_{j})}\simeq20E_{J}$, the gate-charge bias $N_{g}=\frac{C_{g}V_{g}}{(2e)}$, and the constant phase $0\leq\theta_{ex}=\frac{\pi\Phi_{ex}}{\Phi_{0}}\leq\frac{\pi}{2}$. We have also used the quantization, $\hat{\phi}=\Phi_{r}(\hat{a}^{\dag}+\hat{a})$ and $\hat{Q}=iQ_{r}(\hat{a}^{\dag}-\hat{a})$ with $\Phi_{r}=(\frac{\hbar}{2}\sqrt{\frac{L}{C}})^{1/2}$ and $Q_{r}=(\frac{\hbar}{2}\sqrt{\frac{C}{L}})^{1/2}$. $\hat{a}^{\dag}$ and $\hat{a}$ are the creation and annihilation microwave-photon operators. The phase $\theta_{L}$ is given by $\theta_{L}=\frac{\pi\xi\Phi_{r}}{\Phi_{0}}$ with the flux quantum $\Phi_{0}=\frac{\pi\hbar}{e}$. The system operates in the charging limit ($E_{C}\gg E_{J}$). We choose the basis $\{|n_{c}\rangle\otimes|n_{p}\rangle,n_{c}=0,1,2,...;n_{p}=0,1,2,...\}$, where $n_{c}$ denotes the number of excess Cooper pairs in the box while $n_{p}$ corresponds to the intraresonator photon number, to span the Hilbert space. The matrix elements of different operators are given by $\hat{N}=\sum_{n_{c}}n_{c}|n_{c}\rangle\langle n_{c}|$, $\cos\hat{\delta}=\sum_{n_{c}}(|n_{c}\rangle\langle n_{c}+1|+|n_{c}+1\rangle\langle n_{c}|)$, and $\hat{a}=\sum_{n_{p}}\sqrt{n_{p}}|n_{p}-1\rangle\langle n_{p}|$.

The eigenvalues $\mathcal{E}_{\alpha}$ and eigenstates $\Psi_{\alpha}$ of $\hat{H}$, i.e., $\hat{H}\Psi_{\alpha}=\mathcal{E}_{\alpha}\Psi_{\alpha}$, can be derived via the diagonalization method. Figure~\ref{Fig1}b displays $\mathcal{E}_{\alpha}$ vs. $N_{g}$ for different $\theta_{ex}$. It is seen that a number of energy-level avoided crossings occur at the resonant multiphoton interaction. When $\theta_{ex}=0$, a Cooper pair tunneling into (out of) the box leads to the decrement (increment) of the resonator field by $2k$ ($k\in\mathbb{Z}$) photons. For example, the anticrossing, which is labeled as $A_{2}$ and localized at $N^{(A_{2})}_{g}=\frac{1}{2}-\frac{\hbar\omega_{0}}{E_{C}}=0.38$, has a gap of $\hbar\Omega_{A_{2}}=0.51E_{J}$ and happens between two curves associated with $|n_{c}=1\rangle\otimes|n_{p}=0\rangle$ and $|n_{c}=0\rangle\otimes|n_{p}=2\rangle$ at $N_{g}=0$. This is because the operator $\cos[\theta_{L}(\hat{a}^{\dag}+\hat{a})]=\sum_{j}\frac{\theta^{2j}_{L}}{(2j)!}(\hat{a}^{\dag}+\hat{a})^{2j}$ only causes the even-photon transition of the resonator field. In contrast, when $\theta_{ex}=\frac{\pi}{2}$ the photon number varies by $(2k+1)$ for one unit change in the number of Cooper pairs inside the box since the operator $\sin[\theta_{L}(\hat{a}^{\dag}+\hat{a})]=\sum_{j}\frac{\theta^{2j+1}_{L}}{(2j+1)!}(\hat{a}^{\dag}+\hat{a})^{2j+1}$ causes the odd-photon transition. For instances, the anticrossings $A_{1}$ and $A_{3}$ occur at $N^{(A_{1})}_{g}=\frac{1}{2}-\frac{\hbar\omega_{0}}{2E_{C}}=0.44$ and $N^{(A_{3})}_{g}=\frac{1}{2}-\frac{3\hbar\omega_{0}}{2E_{C}}=0.33$ and have energy gaps of $\hbar\Omega_{A_{1}}=0.57E_{J}$ and $\hbar\Omega_{A_{3}}=0.37E_{J}$.

In the following, we focus on the charge qubit formed by $|n_{c}=0,1\rangle$ interacting with the resonator. $N_{g}$ is restricted within the range from $0$ to $0.5$. The operators $\hat{N}$ and $\cos\hat{\delta}$ may be rewritten as $\hat{N}=\frac{1}{2}(1+\hat{\sigma}_{z})$ and $\cos\hat{\delta}=\frac{1}{2}(\hat{\sigma}^{\dag}_{-}+\hat{\sigma}_{-})$ with $\hat{\sigma}_{z}=|n_{c}=1\rangle\langle n_{c}=1|-|n_{c}=0\rangle\langle n_{c}=0|$ and $\hat{\sigma}_{-}=|n_{c}=0\rangle\langle n_{c}=1|$. Taking into consideration of the dissipation effect, the whole system is described by the following master equation~\cite{Book:Carmichael}
\begin{equation}\label{MasterEq}
\textstyle{\frac{d}{dt}\hat{\rho}}=\textstyle{\frac{1}{i\hbar}[\hat{H},\hat{\rho}]+\gamma_{-}{\cal{D}}[\hat{\sigma}_{-}]\hat{\rho}+\frac{\gamma_{\varphi}}{2}{\cal{D}}[\hat{\sigma}_{z}]\hat{\rho}+\kappa{\cal{D}}[\hat{a}]\hat{\rho}},
\end{equation}
where $\hat{\rho}$ is the density operator of the system and we have defined the energy-relaxation and dephasing rates $\gamma_{-}=2\pi\times0.06$ GHz and $\gamma_{\varphi}=2\pi\times0.13$ GHz of the charge qubit~\cite{NJP:Yu2018} and ${\cal{D}}[\hat{o}]\hat{\rho}=\hat{o}\hat{\rho}\hat{o}^{\dag}-\frac{1}{2}\hat{o}^{\dag}\hat{o}\hat{\rho}-\frac{1}{2}\hat{\rho}\hat{o}^{\dag}\hat{o}$. $\kappa$ denotes the photon decay rate and relies on the resonator's quality factor $Q$, i.e., $\kappa=\frac{\omega_{0}}{Q}$. Typically, $Q$ varies from $10^{3}$ to $10^{7}$~\cite{RMP:Xiang2013}, depending on the specific geometric structure of the resonator. Thus, we have $\kappa\ll\gamma_{-},\gamma_{\varphi}$.

\textit{Fock-state maser.} The interface between a two-level atom and a high-$Q$ cavity has been proven to be an ideal platform for realizing the fragile Fock-state radiation via state-reduction~\cite{Nature:Varcoe2000} and trapping-state~\cite{OL:Meystre1988,PRL:Brattke2001} schemes. Recently, Fock states with up to six photons have been also experimentally demonstrated based on a superconducting quantum circuit where a solid-state qubit acts as an intermediary between classical microwave pulses and the resonator field~\cite{Nature:Hofheinz2008,PRL:Wang2008}. Those methods all rely on the single-photon (artificial)-atom-resonator interaction. Here we present an alternative way of producing number-state microwave field by utilizing the resonant nonlinear artificial-atom-resonator coupling.

\begin{figure}
\includegraphics[width=8.5cm]{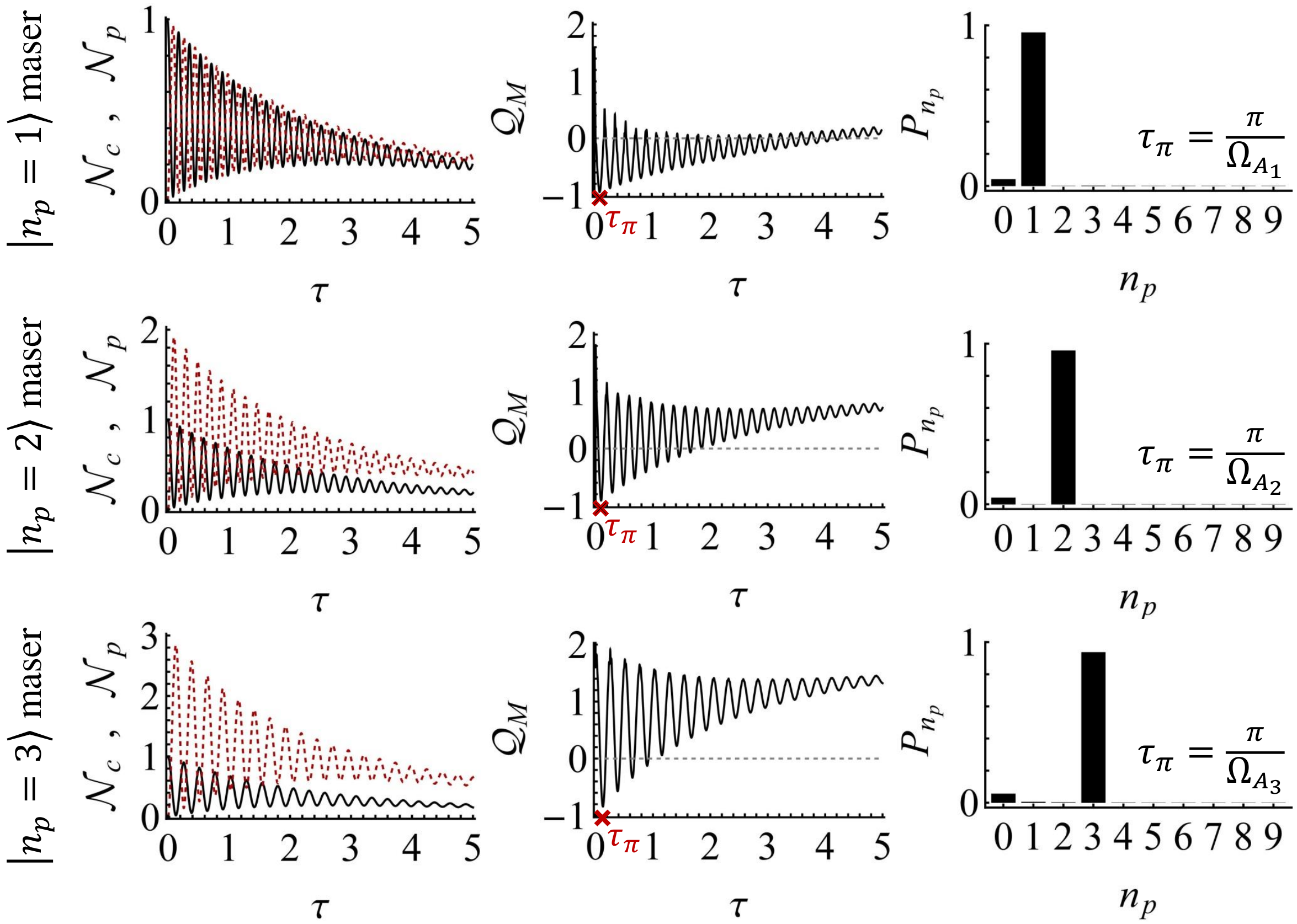}\\
\caption{(Color online) Fock-state maser with $|n_{p}=1,2,3\rangle$. First column: charge-qubit excitation $\mathcal{N}_{c}$ (solid) and intraresonator photon number $\mathcal{N}_{p}$ (dash) vs. pulse length $\tau$. Middle column: dependence of Mandel $\mathcal{Q}_{M}$ parameter (solid) on $\tau$. The dash line corresponds to $\mathcal{Q}_{M}=0$. Last column: photon distribution $P_{n_{p}}$ at $\tau_{\pi}=\frac{\pi}{\Omega_{A_{i=1,2,3}}}$. For all curves, $\xi=1$ and $Q=5\times10^{3}$ which gives $\kappa=2\pi\times4.5$ MHz.}\label{Fig2}
\end{figure}

As examples, we focus on the generation of $|n_{p}=1,2,3\rangle$ resonator field under pulsed operation. The specific implementation may be summarized as follows: ($i$) The system is initially prepared in $|n_{c}=1\rangle\otimes|n_{p}=0\rangle$ and the gate-charge bias $N_{g}$ is set at $0$. The charge qubit hardly interacts with the resonator due to the large detuning at $N_{g}=0$. ($ii$) Then, $N_{g}$ nonadiabatically goes up to $N^{(A_{i=1,2,3})}_{g}$, where the avoided crossing $A_{i=1,2,3}$ occurs [see Fig.~\ref{Fig1}(b)], via tuning the voltage source $V_{g}$. The anticrossing starts the resonant single- or multi-photon ($|n_{c}=1\rangle\otimes|n_{p}=0\rangle-|n_{c}=0\rangle\otimes|n_{p}=1,2,3\rangle$) transition. ($iii$) After a time duration $\tau$, $N_{g}$ is rapidly ramped back to $0$ so as to turn off the qubit-resonator interaction. Indeed, the system works as a pulsed maser, where the artificial qubit plays the role of gain medium.

Solving Eq.~(\ref{MasterEq}) gives us the charge-qubit excitation $\mathcal{N}_{c}=\textrm{Tr}(\hat{\rho}\hat{N})$ and the intraresonator photon number $\mathcal{N}_{p}=\textrm{Tr}(\hat{\rho}\hat{a}^{\dag}\hat{a})$ at the end of ramp pulse. Figure~\ref{Fig2} depicts the dependence of $\mathcal{N}_{c,p}$ on the pulse length $\tau$. It is seen that the Rabi oscillation of the maser between $|n_{p}=0\rangle$ and $|n_{p}=1,2,3\rangle$ is strongly damped because of large $\gamma_{-}$ and $\gamma_{\varphi}$. The Mandel $\mathcal{Q}_{M}=\textstyle{[\textrm{Tr}(\hat{\rho}\hat{a}^{\dag}\hat{a}\hat{a}^{\dag}\hat{a})-\mathcal{N}_{p}^{2}]/\mathcal{N}_{p}}$ parameter is commonly employed to measure the departure of a radiation from the classical field. $\mathcal{Q}_{M}<0$ corresponds to a sub-Poissonian (nonclassical) photon-number statistics and $\mathcal{Q}_{M}$ arrives at $-1$ for Fock states. As illustrated in Fig.~\ref{Fig2}, $\mathcal{Q}_{M}$ minimizes at the $\pi$-pulse length $\tau_{\pi}=\frac{\pi}{\Omega_{A_{i=1,2,3}}}$. However, due to the charge qubit's relaxation $\mathcal{Q}_{M}$ does not reach $-1$. Nevertheless, at $\tau_{\pi}$ the photon distribution $P_{n_{p}}=\langle n_{p}|\textrm{Tr}_{c}(\hat{\rho})|n_{p}\rangle$ of the maser field, where $\textrm{Tr}_{c}$ denotes the trace over the charge-qubit coordinates, is maximized at the corresponding Fock state with a probability over 90~\%. Thus, the production of Fock-state pulsed maser requires the ramp-pulse length $\tau=\tau_{\pi}$. The lifetime of the Fock-state field is limited by $\kappa^{-1}$. The approach used here to produce multiphoton Fock-state radiation is applicable to arbitrary photon numbers smaller than $n^{max}_{p}=\frac{E_{C}}{\hbar\omega_{0}}\simeq8$, higher than which the populations in $|n_{c}=1\rangle\otimes|n_{p}\neq0\rangle$ rise and the fidelity of Fock-state generation degrades.

\textit{Radiation squeezing.} The nonlinear qubit-resonator interface may be also applied to generate the squeezed resonator field. In general, the two-photon parametric process has been envisaged as a source of squeezing the radiation, for which we set $\theta_{ex}=0$. Further, in the weak-coupling limit ($\theta_{L}\sim0$ with $\xi\sim0$), we approximate $\cos[\theta_{L}(\hat{a}^{\dag}+\hat{a})]$ with the Taylor series expansion of $\theta_{L}$ up to the second order, i.e., $\cos[\theta_{L}(\hat{a}^{\dag}+\hat{a})]\simeq1-\frac{\theta^{2}_{L}}{2}(\hat{a}^{\dag}+\hat{a})^{2}$. Thus, the Hamiltonian is simplified as $\hat{H}\simeq\hat{H}_{c}+\hat{H}_{p}$
with the free charge-qubit Hamiltonian
\begin{equation}
\hat{H}_{c}=\textstyle{E_{C}(\frac{1}{2}-N_{g})\hat{\sigma}_{z}-\frac{E_{J}}{2}(\hat{\sigma}^{\dag}_{-}+\hat{\sigma}_{-})}.
\end{equation}
and the Hamiltonian associated with the resonator field
\begin{equation}
\hat{H}_{p}=\textstyle{\hbar\omega_{0}\hat{a}^{\dag}\hat{a}+\frac{\theta^{2}_{L}E_{J}}{4}(\hat{a}^{\dag}+\hat{a})^{2}}(\hat{\sigma}^{\dag}_{-}+\hat{\sigma}_{-}).
\end{equation}
To resonantly drive the two-photon transition, we sweep $N_{g}$ periodically with an amplitude $\Delta N_{g}=\frac{\hbar\Omega}{2E_{C}}$ ($\Omega\ll\omega_{0}$) and a rate $(2\omega_{0})$ around $N^{(A_{2})}_{g}$. Hence, we have $E_{C}(\frac{1}{2}-N_{g})=2\hbar\omega_{0}+\hbar\Omega\cos2\omega_{0}t$. It is unpractical to numerically simulate the time evolution of the whole system directly based on the master equation~(\ref{MasterEq}). As we will see below, in the adiabatic limit the charge-qubit and resonate-field dynamics can be approximately treat separately, much simplifying the calculation.

Due to $\kappa\ll\gamma_{-},\gamma_{\varphi}$, the charge qubit arrives at the quasi steady state much faster than the resonator field. In addition, the reaction of the qubit-resonator coupling on the charge-qubit dynamics is weak because of $\theta^{2}_{L}\sim0$. Thus, in the adiabatic approximation~\cite{BOOK:Scully} we consider the dissipative time evolution of the charge qubit separately, which is governed by the following master equation
\begin{equation}\label{MasterEq2}
\textstyle{\frac{d}{dt}\hat{\rho}^{(c)}}=\textstyle{\frac{1}{i\hbar}[\hat{H}_{c},\hat{\rho}^{(c)}]+\gamma_{-}{\cal{D}}[\hat{\sigma}_{-}]\hat{\rho}^{(c)}+\frac{\gamma_{\varphi}}{2}{\cal{D}}[\hat{\sigma}_{z}]\hat{\rho}^{(c)}}.
\end{equation}
$\hat{\rho}^{(c)}$ is the charge-qubit density matrix operator. Equation~(\ref{MasterEq2}) gives us
\begin{subequations}
\begin{eqnarray}
\textstyle{\frac{d}{dt}\rho^{(c)}_{11}}&=&\textstyle{-\gamma_{-}\rho^{(c)}_{11}+i\frac{E_{J}}{2\hbar}[(\rho^{(c)}_{10})^{\ast}-\rho^{(c)}_{10}]},\\
\nonumber\textstyle{\frac{d}{dt}\rho^{(c)}_{10}}&=&\textstyle{(-\frac{\gamma_{-}}{2}-\gamma_{\varphi}-i2\omega_{0}-i\Omega\cos2\omega_{0}t)\rho^{(c)}_{10}}\\
&&\textstyle{+i\frac{E_{J}}{2\hbar}(1-2\rho^{(c)}_{11})},
\end{eqnarray}
\end{subequations}
with $\rho^{(c)}_{uv}=\langle n_{c}=u|\hat{\rho}^{(c)}|n_{c}=v\rangle$ ($u,v=0,1$). Substituting $\rho^{(c)}_{10}=\tilde{\rho}^{(c)}_{10}e^{-i2\omega_{0}t-i\frac{\Omega}{2\omega_{0}}\sin2\omega_{0}t}$ into above two equations and applying the rotating-wave approximation (RWA), we arrive at
\begin{subequations}
\begin{eqnarray}
\textstyle{\frac{d}{dt}\rho^{(c)}_{11}}&\simeq&\textstyle{-\gamma_{-}\rho^{(c)}_{11}-i\frac{\Omega}{2\omega_{0}}\frac{E_{J}}{4\hbar}[(\tilde{\rho}^{(c)}_{10})^{\ast}-\tilde{\rho}^{(c)}_{10}]},\\
\textstyle{\frac{d}{dt}\tilde{\rho}^{(c)}_{10}}&\simeq&\textstyle{(-\frac{\gamma_{-}}{2}-\gamma_{\varphi})\tilde{\rho}^{(c)}_{10}-i\frac{\Omega}{2\omega_{0}}\frac{E_{J}}{4\hbar}(1-2\rho^{(c)}_{11})}.
\end{eqnarray}
\end{subequations}
The steady-state (ss) solutions are derived as $\rho^{(c,ss)}_{11}=2\lambda$ and $\tilde{\rho}^{(c,ss)}_{10}=-i\lambda$ with $\lambda=\textstyle{\gamma_{-}(\frac{\Omega}{2\omega_{0}}\frac{E_{J}}{4\hbar})/[\gamma_{-}(\frac{\gamma_{-}}{2}+\gamma_{\varphi})+4(\frac{\Omega}{2\omega_{0}}\frac{E_{J}}{4\hbar})^{2}]}$. In the limit of $\Omega\ll(2\omega_{0})$, we have $\lambda\ll1$ and the charge qubit is mostly populated in the ground state $|n_{c}=0\rangle$.

Since the charge qubit is in the quasi steady state, we replace $\hat{\sigma}_{-}$ in $\hat{H}_{p}$ by $\tilde{\rho}^{(c,ss)}_{10}e^{-i2\omega_{0}t-i\frac{\Omega}{2\omega_{0}}\sin2\omega_{0}t}$. Using the master equation for the resonator-field density matrix operator $\hat{\rho}^{(p)}$,
\begin{equation}\label{MasterEq3}
\textstyle{\frac{d}{dt}\hat{\rho}^{(p)}}=\textstyle{\frac{1}{i\hbar}[\hat{H}_{p},\hat{\rho}^{(p)}]+\kappa{\cal{D}}[\hat{a}]\hat{\rho}^{(p)}},
\end{equation}
one obtains
\begin{subequations}
\begin{eqnarray}
\textstyle{\frac{d}{dt}\mathcal{A}}&=&\textstyle{(-\frac{\kappa}{2}-i\omega_{0})\mathcal{A}+i\lambda\theta^{2}_{L}\frac{E_{J}}{\hbar}(\mathcal{A}^{\ast}+\mathcal{A})\sin2\omega_{0}t},~~~~~\\
\nonumber\textstyle{\frac{d}{dt}\mathcal{B}}&=&\textstyle{(-\kappa-i2\omega_{0}+i2\lambda\theta^{2}_{L}\frac{E_{J}}{\hbar}\sin2\omega_{0}t)\mathcal{B}}\\
&&\textstyle{+i\lambda\theta^{2}_{L}\frac{E_{J}}{\hbar}(2\mathcal{N}_{p}+1)\sin2\omega_{0}t},\\
\textstyle{\frac{d}{dt}\mathcal{N}_{p}}&=&\textstyle{-\kappa\mathcal{N}_{p}+i\lambda\theta^{2}_{L}\frac{E_{J}}{\hbar}(\mathcal{B}^{\ast}-\mathcal{B})\sin2\omega_{0}t}.
\end{eqnarray}
\end{subequations}
Here we have defined $\mathcal{A}=\textrm{Tr}(\hat{\rho}^{(p)}\hat{a})$ and $\mathcal{B}=\textrm{Tr}(\hat{\rho}^{(p)}\hat{a}\hat{a})$ and used $\mathcal{N}_{p}=\textrm{Tr}(\hat{\rho}^{(p)}\hat{a}^{\dag}\hat{a})$. Substituting $\mathcal{A}=\tilde{\mathcal{A}}e^{-i\omega_{0}t}$ and $\mathcal{B}=\tilde{\mathcal{B}}e^{-i2\omega_{0}t-i\frac{\lambda\theta^{2}_{L}E_{J}}{\hbar\omega_{0}}\cos2\omega_{0}t}$ into above equations and applying the RWA, we arrive at
\begin{subequations}
\begin{eqnarray}
\textstyle{\frac{d}{dt}\tilde{\mathcal{A}}}&\simeq&\textstyle{-\frac{\kappa}{2}\tilde{\mathcal{A}}-\frac{\lambda\theta^{2}_{L}}{2}\frac{E_{J}}{\hbar}\tilde{\mathcal{A}}^{\ast}},\\
\nonumber\textstyle{\frac{d}{dt}\tilde{\mathcal{B}}}&\simeq&\textstyle{-\kappa\tilde{\mathcal{B}}-\frac{\lambda\theta^{2}_{L}}{2}\frac{E_{J}}{\hbar}(2\mathcal{N}_{p}+1)},\\
\textstyle{\frac{d}{dt}\mathcal{N}_{p}}&\simeq&\textstyle{-\kappa\mathcal{N}_{p}-\frac{\lambda\theta^{2}_{L}}{2}\frac{E_{J}}{\hbar}(\tilde{\mathcal{B}}^{\ast}+\tilde{\mathcal{B}})}.
\end{eqnarray}
\end{subequations}
The steady-state solutions are then given by $\tilde{\mathcal{A}}^{(ss)}=0$, $\tilde{\mathcal{B}}^{(ss)}=-\frac{\mu}{2}\frac{1}{1-\mu^{2}}$, and $\mathcal{N}^{(ss)}_{p}=\frac{1}{2}\frac{\mu^{2}}{1-\mu^{2}}$ with $\mu=\frac{\lambda\theta^{2}_{L}E_{J}}{\hbar\kappa}$. Despite $\lambda\ll1$ and $\theta_{L}\sim0$, reducing $\kappa$ may enhance $\mu$. As $\mu$ approaches unity, the intraresonator photon number $\mathcal{N}^{(ss)}_{p}$ is strongly increased. However, the charge-qubit picture becomes invalid when $\mathcal{N}^{(ss)}_{p}>n^{max}_{p}$ since the number of excess Cooper pairs in the box exceeds unity. Thus, the maximum of $\mu$ is given by $\mu^{max}=\sqrt{\frac{2n^{max}_{p}}{1+2n^{max}_{p}}}\simeq0.97$, leading to the required $\kappa=2\pi\times0.1$ MHz with the achievable quality factor $Q\sim10^{5}$.

We then define two quadrature operators $\hat{X}_{1}=\frac{1}{2}(\hat{a}e^{i\omega_{0}t}+\hat{a}^{\dag}e^{-i\omega_{0}t})$ and $\hat{X}_{2}=\frac{1}{2i}(\hat{a}e^{i\omega_{0}t}-\hat{a}^{\dag}e^{-i\omega_{0}t})$ for the resonator field. In the steady state, the variances $\Delta X^{(ss)}_{i=1,2}=\sqrt{\langle\hat{X}^{2}_{i}\rangle^{(ss)}-(\langle\hat{X}_{i}\rangle^{(ss)})^{2}}$ are derived as $\Delta X^{(ss)}_{1}\simeq\frac{1}{2\sqrt{1+u}}$ and $\Delta X^{(ss)}_{2}\simeq\frac{1}{2\sqrt{1-u}}$. The best squeezing is achieved when $\mu=\mu^{max}$, i.e., $\Delta X^{(ss)}_{1}\simeq\frac{1}{2\sqrt{1+\mu^{max}}}=0.36$. Further reducing $\Delta X^{(ss)}_{1}$ requires a larger $n^{max}_{p}$ (i.e., a larger $E_{C}$ and a smaller $\omega_{0}$) and a smaller $\kappa$ (i.e., a higher $Q$).

\begin{figure}
\includegraphics[width=8.5cm]{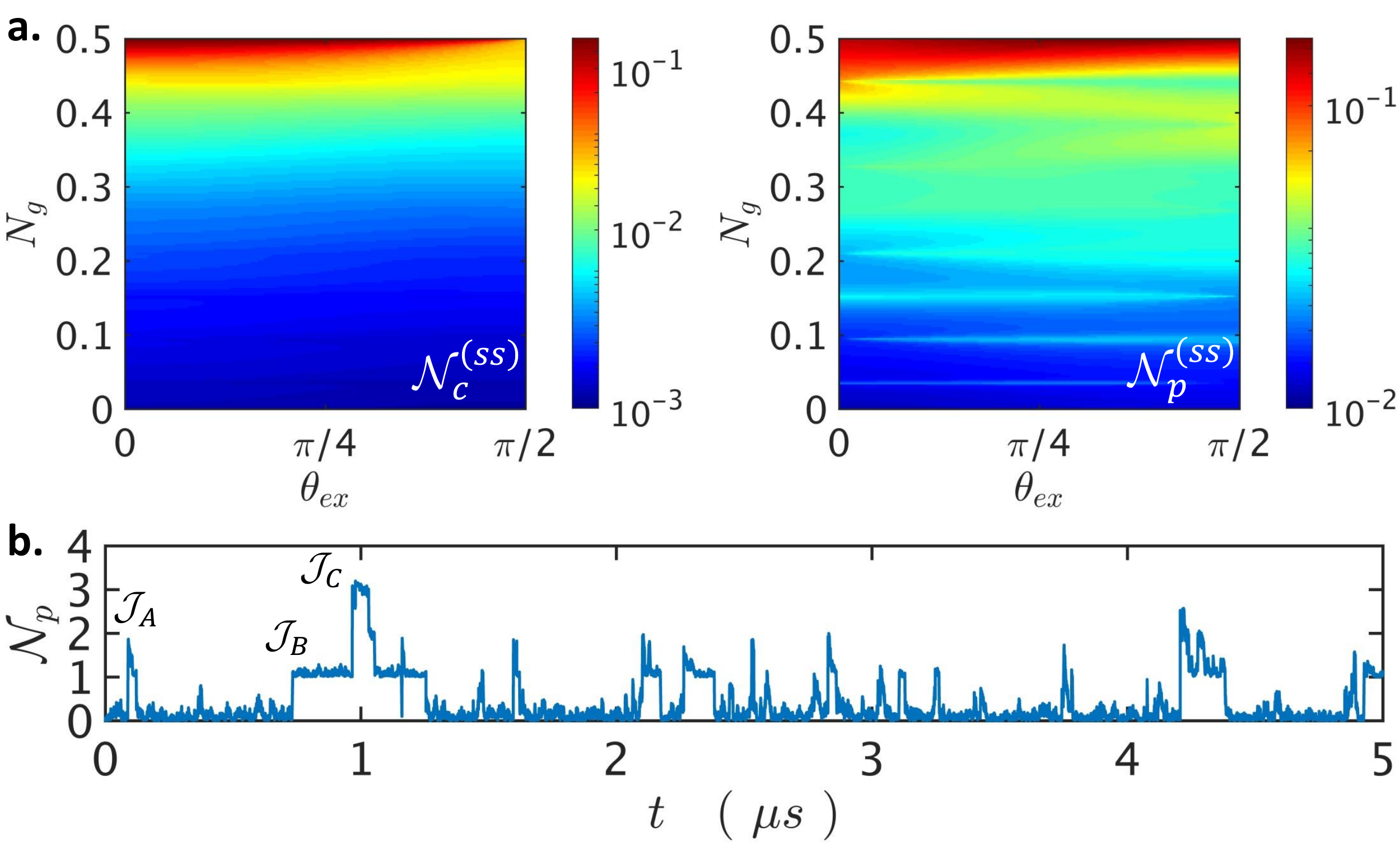}\\
\caption{(Color online) (a) Steady-state $\mathcal{N}^{(ss)}_{c,p}$ as a function of $N_{g}$ and $\theta_{ex}$. (b) Trajectory of photon-number expectation value $\mathcal{N}_{p}$ with $N_{g}=\frac{1}{2}$ and $\theta_{ex}=0$. The system is initially prepared in $|n_{c}=0\rangle\otimes|n_{p}=0\rangle$. All system parameters are same to Fig.~\ref{Fig2}.}\label{Fig3}
\end{figure}

\textit{Multiphoton quantum jumps.} The nonlinear interacting system may arrive at a nontrivial steady state ($\mathcal{N}^{(ss)}_{c,p}\neq0$) even without an external driving. Setting $\frac{d}{dt}\hat{\rho}=0$ in Eq.~(\ref{MasterEq}), one can derive the steady-state density matrix of the system, i.e., $\hat{\rho}(t\rightarrow\infty)$, by employing the diagonalization approach~\cite{JOSAB:Yu}. The results are shown in~\ref{Fig3}(a). It is seen that both $\mathcal{N}^{(ss)}_{c}$ and $\mathcal{N}^{(ss)}_{p}$ are maximized at the sweet spot $N_{g}=0.5$ because the Cooper-pair tunneling resonantly pumps the qubit from $|0\rangle$ to $|1\rangle$. In addition, due to $E_{J}\sim\hbar\omega_{0}$, the qubit-resonator interaction reaches the ultrastrong-coupling regime and the counter-rotating terms also play an important role in the intraresonator power enhancement.

We further employ the Monte Carlo wave-function method \cite {JOSAB:Molmer} to look into more detail about the dissipative artificial-atom-photon interaction in the steady state. Figure \ref{Fig3}(b) illustrates a quantum trajectory corresponding to the time evolution of the photon-number expectation value $\mathcal{N}_{p}$ at $N_{g}=\frac{1}{2}$ and $\theta_{ex}=0$. As one can see, multiple quantum jumps are presented and most of them are from the charge-qubit spontaneous decaying because of $\kappa\ll\gamma_{-},\gamma_{\varphi}$. Such quantum-jump dynamics have been experimentally observed in macroscopic superconducting quantum systems~\cite{PRL:Yu2008,PRL:Vijay2011,PRL:Vool2014}.

At any time the system is in a superposition state completely composed of odd or even Fock states. After each photonic jump associated with the jump operator $\sqrt{\kappa}\hat{a}$, odd (even) Fock states transfer to lower even (odd) Fock states. Figure \ref{Fig3}(b) clearly shows a number of sharp jumps with the photon-number differences of about one and two (or more than two). Some of them are induced by $\sqrt{\kappa}\hat{a}$ [e.g., the jumps $\mathcal{J}_{A,B}$ in Fig.~\ref{Fig3}(b)] while others are caused by the qubit decay associated with the jump operator $\sqrt{\gamma_{-}}\hat{\sigma}_{-}$ [e.g., the jump $\mathcal{J}_{C}$ in Fig.~\ref{Fig3}(b)]. We consider the trajectory period from $\mathcal{J}_{A}$ to $\mathcal{J}_{C}$. Before $\mathcal{J}_{A}$ occurs, the system is mainly in the vacuum $|n_{p}=0\rangle$ state with the secondary-weighted component of $|n_{p}=2\rangle$. Subsequently, the jump $\mathcal{J}_{A}$ collapses the system's wave function onto $|n_{p}=1\rangle$. Similarly, at $\mathcal{J}_{B}$ the system collapses onto $|n_{c}=0\rangle\otimes|n_{p}=1\rangle$. Afterwards, the system evolves to a superposition state with a component of $|n_{c}=1\rangle\otimes|n_{p}=3\rangle$ via the counter rotating parametric process (i.e., $\hat{\sigma}^{\dag}_{-}\hat{a}^{\dag}\hat{a}^{\dag}$). Then, the jump operator $\sqrt{\gamma_{-}}\sigma_{-}$ collapses the system's wave function onto $|n_{c}=0\rangle\otimes|n_{p}=3\rangle$ at $\mathcal{J}_{C}$.

\textit{Conclusion.} We have studied a nonlinear circuit-QED scheme where a charge qubit is inductively coupled to a resonator via multiphoton processes. Such an architecture can produce arbitrary photon-number states of microwave field with high fidelity, which is of particular importance for the linear photonic quantum computing~\cite{Nature:Knill2001,RMP:Kok2007}. The squeezed radiation may be generated by the parametric qubit-resonator interface, potentially applicable to the quantum-limited microwave amplifier~\cite{NatPhys:CastellanosBeltran2008}. This platform can be also employed to investigate fundamental principle of multiphoton light-matter interaction in the ultrastrong-coupling regime~\cite{PRA:Garziano2015}, where, as we have illustrated, the counter-rotating terms and qubit decay jointly lead to upwards multiphoton quantum jumps of resonator field. The rapid development of cryogenic electronics may sustain much longer lifetime of superconducting circuits in the future, enabling more potential applications of circuit QED in quantum information processing and fundamental new physics.

\begin{acknowledgments}
This research has been supported by the National Research Foundation Singapore \& by the Ministry of Education Singapore Academic Research Fund Tier 2 (Grant No. MOE2015-T2-1-101).
\end{acknowledgments}

\end{document}